\documentclass[twocolumn,prl]{revtex4}
\usepackage{epsfig,amsmath,amssymb,graphics,color,calc}

\begin{document}

\title{Inhomogeneous Mode-Coupling Theory and Growing Dynamic Length in Supercooled Liquids}
\author{Giulio Biroli$^{1}$, Jean-Philippe
Bouchaud$^{2,3}$, Kunimasa Miyazaki$^{4}$, David R. Reichman$^{4}$}
\affiliation{
$^1$ Service de Physique Th{\'e}orique,
Orme des Merisiers -- CEA Saclay, 91191 Gif sur Yvette Cedex, France.}
\affiliation{
$^2$ Service de Physique de l'{\'E}tat Condens{\'e},
Orme des Merisiers -- CEA Saclay, 91191 Gif sur Yvette Cedex, France.}
\affiliation{
$^3$ Science \& Finance, Capital Fund Management, 6 Bd
Haussmann, 75009 Paris, France.}
\affiliation{
$^4$ Department of Chemistry, Columbia University, 3000 Broadway, New York, NY 10027, USA.}


\newcommand \be  {\begin{equation}}
\newcommand \bea {\begin{eqnarray} \nonumber }
\newcommand \ee  {\end{equation}}
\newcommand \eea {\end{eqnarray}}
\newcommand{\siml}{\stackrel{<}{\sim}}
\newcommand{\subs}[1
]{{\mbox{\scriptsize #1}}}
\newcommand{\bgammadot}{\dot{\mbox{\boldmath$\gamma$}}}
\newcommand{\gammadot}{\dot{\gamma}}
\newcommand{\gne}{g_{\mbox{\scriptsize n.e}} }
\newcommand{\gslow}{g_{\mbox{\scriptsize mct}} }
\newcommand{\lgle}{\left\langle}
\newcommand{\rgle}{\right\rangle}
\newcommand{\diff}[2]{\frac{\mbox{d} #1}{\mbox{d} #2 } }
\newcommand{\pdif}[2]{\frac{\partial #1}{\partial #2 } }
\newcommand{\ppdif}[3]{\frac{\partial^2 #1}{\partial #2\partial #3} }
\newcommand{\calC}{{\cal C}}
\newcommand{\calD}{{\cal D}}
\newcommand{\calF}{{\cal F}}
\newcommand{\calG}{{\cal G}}
\newcommand{\calH}{{\cal H}}
\newcommand{\calM}{{\cal M}}
\newcommand{\calL}{{\cal L}}
\newcommand{\calO}{{\cal O}}
\newcommand{\calP}{{\cal P}}
\newcommand{\calQ}{{\cal Q}}
\newcommand{\calR}{{\cal R}}
\newcommand{\calS}{{\cal S}}
\newcommand{\calT}{{\cal T}}
\newcommand{\calU}{{\cal U}}
\newcommand{\calV}{{\cal V}}
\newcommand{\calMo}{\left.{\cal M}_{1}\right.}
\newcommand{\calLo}{\left.{\cal L}_{1}\right.}
\newcommand{\calMoa}{\left.{\cal M}_{1}^{(a)}\right.}
\newcommand{\calMob}{\left.{\cal M}_{1}^{(b)}\right.}
\newcommand{\calLoa}{\left.{\cal L}_{1}^{(a)}\right.}
\newcommand{\calLob}{\left.{\cal L}_{1}^{(b)}\right.}
\newcommand{\calMt}{\left.{\cal M}_{2}\right.}
\newcommand{\calLt}{\left.{\cal L}_{2}\right.}
\newcommand{\calMta}{\left.{\cal M}_{2}^{(a)}\right.}
\newcommand{\calMtb}{\left.{\cal M}_{2}^{(b)}\right.}
\newcommand{\calMtc}{\left.{\cal M}_{2}^{(c)}\right.}
\newcommand{\calLta}{\left.{\cal L}_{2}^{(a)}\right.}
\newcommand{\calLtb}{\left.{\cal L}_{2}^{(b)}\right.}
\newcommand{\calLtc}{\left.{\cal L}_{2}^{(c)}\right.}
\newcommand{\cm}{\mbox{cm}}
\newcommand{\dd}{{\mbox{d}}}
\newcommand{\delf}[2]{\frac{\delta #1}{\delta #2 } }
\newcommand{\Dt}{{\Delta t}}
\newcommand{\dQ}{{\delta Q}}
\newcommand{\dx}{{\delta x}}
\newcommand{\e}{\mbox{\large e}}
\newcommand{\tk}{\tilde{k}}
\newcommand{\bCz}{\left.{\bf C}_{0}\right.}
\newcommand{\bGz}{\left.{\bf G}_{0}\right.}
\newcommand{\bLz}{\left.{\bf L}_{0}\right.}
\newcommand{\bMz}{\left.{\bf M}_{0}\right.}
\newcommand{\Cz}{\left.C_{0}\right.}
\newcommand{\Gz}{\left.G_{0}\right.}
\newcommand{\Lz}{\left.L_{0}\right.}
\newcommand{\mct}{\mbox{\scriptsize mct}}
\newcommand{\Mz}{\left.M_{0}\right.}
\newcommand{\bGammaz}{\left.{\mbox{\boldmath$\Gamma$}}_{0} \right.}
\newcommand{\bchiz}{\left.{\mbox{\boldmath$\chi$}}_{0} \right.}
\newcommand{\bbetaz}{\left.{\mbox{\boldmath$\eta$}}_{0} \right.}
\newcommand{\Gammaz}{\left.\Gamma_{0} \right.}
\newcommand{\chiz}{\left.\chi_{0} \right.}
\newcommand{\etaz}{\left.\eta_{0}\right.}
\newcommand{\hGz}{{\left.\hat{{\bf G}}_{0}\right.}}
\newcommand{\hhGz}{\left.{\hat{G}_{0}}\right.}
\newcommand{\hG}{\hat{{\bf G}}}
\newcommand{\hhG}{\hat{G}}
\newcommand{\hC}{\hat{C}}
\newcommand{\ba}{{\bf a}}
\newcommand{\bff}{{\bf f}}
\newcommand{\bg}{{\bf g}}
\newcommand{\bh}{{\bf h}}
\newcommand{\bj}{{\bf j}}
\newcommand{\bk}{{\bf k}}
\newcommand{\bp}{{\bf p}}
\newcommand{\bq}{{\bf q}}
\newcommand{\br}{{\bf r}}
\newcommand{\bv}{{\bf v}}
\newcommand{\bx}{{\bf x}}
\newcommand{\by}{{\bf y}}
\newcommand{\bz}{{\bf z}}
\newcommand{\bA}{{\bf A}}
\newcommand{\bC}{{\bf C}}
\newcommand{\bD}{{\bf D}}
\newcommand{\bF}{{\bf F}}
\newcommand{\bG}{{\bf G}}
\newcommand{\bJ}{{\bf J}}
\newcommand{\bK}{{\mbox{\boldmath$K$}}}
\newcommand{\bL}{{\mbox{\boldmath$L$}}}
\newcommand{\bM}{{\mbox{\boldmath$M$}}}
\newcommand{\bP}{{\bf P}}
\newcommand{\bR}{{\bf R}}
\newcommand{\bS}{{\bf S}}
\newcommand{\bT}{{\bf T}}
\newcommand{\bU}{{\bf U}}
\newcommand{\bV}{{\bf V}}
\newcommand{\bW}{{\bf W}}
\newcommand{\bX}{{\mbox{\boldmath$X$}}}
\newcommand{\bY}{{\bf Y}}
\newcommand{\bZ}{{\bf Z}}
\newcommand{\bbeta}{{\mbox{\boldmath$\eta$}}}
\newcommand{\beps}{{\mbox{\boldmath$\varepsilon$}}}
\newcommand{\bgamma}{{\mbox{\boldmath$\gamma$}}}
\newcommand{\bchi}{{\mbox{\boldmath$\chi$}}}
\newcommand{\bmu}{{\mbox{\boldmath$\mu$}}}
\newcommand{\bomega}{{\mbox{\boldmath$\omega$}}}
\newcommand{\bphi}{{\mbox{\boldmath$\phi$}}}
\newcommand{\bsigma}{{\mbox{\boldmath$\sigma$}}}
\newcommand{\bxi}{{\mbox{\boldmath$\xi$}}}
\newcommand{\bGamma}{{\mbox{\boldmath$\Gamma$}}}
\newcommand{\bLambda}{{\mbox{\boldmath$\Lambda$}}}
\newcommand{\bOmega}{{\mbox{\boldmath$\Omega$}}}
\newcommand{\bXi}{{\mbox{\boldmath$\Xi$}}}
\newcommand{\bSigma}{{\mbox{\boldmath$\Sigma$}}}
\newcommand{\drho}{{\delta\rho}}
\newcommand{\hA}{\hat{A}}
\newcommand{\hB}{\hat{B}}
\newcommand{\he}{\hat{\bf e}}
\newcommand{\hF}{\hat{\bf F}}
\newcommand{\hhF}{\hat{F}}
\newcommand{\hH}{\hat{H}}
\newcommand{\hk}{\hat{\bf k}}
\newcommand{\hK}{\hat{{\mbox{\boldmath$K$}}}}
\newcommand{\hhK}{\hat{K}}
\newcommand{\hJ}{\hat{\bf J}}
\newcommand{\hhJ}{\hat{J}}
\newcommand{\hL}{\hat{L}}
\newcommand{\hM}{\hat{{\mbox{\boldmath$M$}}}}
\newcommand{\hhM}{\hat{M}}
\newcommand{\hq}{\hat{\bf q}}
\newcommand{\hR}{\hat{R}}
\newcommand{\hr}{\hat{\bf r}}
\newcommand{\hhk}{\hat{k}}
\newcommand{\hqq}{\hat{q}}
\newcommand{\hrr}{\hat{r}}
\newcommand{\hS}{\hat{\bf S}}
\newcommand{\hhS}{\hat{S}}
\newcommand{\hU}{\hat{U}}
\newcommand{\hu}{\hat{u}}
\newcommand{\hV}{\hat{\bf V}}
\newcommand{\hhV}{\hat{V}}
\newcommand{\hW}{\hat{W}}
\newcommand{\hx}{\hat{\bx}}
\newcommand{\hhx}{\hat{x}}
\newcommand{\hy}{\hat{y}}
\newcommand{\hGamma}{\hat{\Gamma}}
\newcommand{\hOmega}{\hat{\mbox{\boldmath$\Omega$}}}
\newcommand{\hhOmega}{\hat{\Omega}}
\newcommand{\hrho}{\hat{\rho}}
\newcommand{\hxi}{\hat{\mbox{\boldmath$\xi$}}}
\newcommand{\hhtau}{\hat{\tau}}
\newcommand{\hhxi}{\hat{\xi}}
\newcommand{\kb}{k_{\mbox{\scriptsize B}}}
\newcommand{\eq}{\mbox{\scriptsize eq}}
\newcommand{\HK}{\mbox{\scriptsize HK}}
\newcommand{\ideal}{\mbox{\scriptsize ideal}}
\newcommand{\inter}{\mbox{\scriptsize int}}
\newcommand{\rev}{\mbox{\scriptsize rev}}
\newcommand{\tot}{\mbox{\scriptsize tot}}
\newcommand{\intt}[1]{{\int_{-\infty}^{\infty}\!\!\!\!\dd t_{#1}}}
\newcommand{\num}[1]{\rule{0mm}{1.0ex} {\textstyle #1}}
\newcommand{\den}[1]{\rule{0mm}{1.8ex} {\textstyle #1}}
\newcommand{\matfrac}[2]{\frac{\num{#1}}{\den{#2}}}
\newcommand{\dens}{n~\!\!}
\newcommand{\hdens}{\hat{\dens}}
\newcommand{\bfR}{{\mbox{\boldmath$f$}}}
\newcommand{\fR}{f}
\newcommand{\fRz}{\left.f_{0}\right.}
\newcommand{\bfRz}{\left.{\bf f}_{0}\right.}
\newcommand{\dlagle}{\left\langle \!\left\langle}
\newcommand{\dragle}{\right\rangle\!\right\rangle}
\newcommand{\X}[1]{\dlagle X(#1) \dragle}
\newcommand{\bKz}{\left.{\mbox{\boldmath$K$}}_{0} \right.}
\newcommand{\K}{{K}}
\newcommand{\Kz}{\left.{K}_{0} \right.}
\newcommand{\hSthree}{\hat{S}^{(3)}}
\newcommand{\hKzero}{\hat{K}^{(0)}}
\newcommand{\E}{E}
\newcommand{\an}[1]{a_{#1}}
\newcommand{\cn}[1]{a_{#1}^{\dagger}}
\newcommand{\bra}[1]{\left\langle {#1} \right\rvert}
\newcommand{\ket}[1]{\left\lvert  {#1} \right\rangle}
\newcommand{\bracket}[2]{\left\langle {#1} | {#2}\right\rangle}
\newcommand{\qave}[3]{\left\langle {#1} 
\left\lvert {#2} \right\rvert {#3}\right\rangle}
\newcommand{\ave}[1]{\left\langle {#1}  \right\rangle}
\newcommand{\dave}[1]{\left\langle \!\left\langle{#1}\right\rangle\!\right\rangle}
\newcommand{\Sne}{S_{\mbox{\scriptsize n.e}} }
\newcommand{\Tf}{{\cal T}_{\leftarrow}}
\newcommand{\Tb}{{\cal T}_{\rightarrow}}
\newcommand{\un}[1]{{\underline{#1}}}

\begin{abstract}
We extend Mode-Coupling Theory (MCT) to inhomogeneous situations, relevant for  
supercooled liquid in pores, close to a surface, or in an external field. We compute the response of the 
dynamical structure factor to a static inhomogeneous external potential and provide the first direct evidence that
the standard formulation of MCT is associated with a diverging length scale. We find in particular that 
the so called ``cages'' are in fact extended objects. Although close to the transition the dynamic length grows as $|T-T_{c}|^{-1/4}$ 
in {\it both} the $\beta$ and $\alpha$ regimes, our results suggest that the fractal dimension of correlated clusters 
is larger in the $\alpha$ regime. We also derive inhomogeneous MCT equations valid to second order in gradients.
\end{abstract}

\maketitle
It is becoming increasingly clear that the viscous slowing down of supercooled liquids, jammed colloids
or granular assemblies is accompanied by a growing dynamic length scale, whereas all static correlation functions
remain short-ranged. This somewhat unusual scenario, suggested by the experimental discovery of strong
dynamical heterogeneities in glass-formers \cite{Ediger}, has been substantiated by detailed numerical simulations 
\cite{Harrowell,YO,Parisilength,Glotzer,Berthier}, explicit solution of simplified models \cite{SR,GC} and very recent direct 
experiments \cite{MDB,Science} where four-point spatio-temporal correlators are measured. From a theoretical point of view, 
our understanding of supercooled liquids owes much to the Mode-Coupling Theory (MCT) of the glass transition. Although
approximate in
nature, MCT has achieved many qualitative and quantitative successes in explaining various experimental and numerical 
results \cite{Gotze1,Gotze2}. In spite of early insights \cite{KW}, the freezing predicted by MCT was repeatedly argued to be a small scale caging phenomenon, 
without any diverging collective length scale. This, however, is rather surprising from a physical point of view, 
since one expects on general grounds that a diverging relaxation time should involve an infinite number 
of particles \cite{ft0}. Building upon the important work of Franz and Parisi \cite{FP}, two of us (BB) \cite{EPL}
suggested a way to reconcile MCT with physical intuition. Within a field theory formulation of MCT, BB showed that the 
four-point density correlation function is given by the so-called `ladder' diagrams that 
indeed lead, upon resummation, to a diverging dynamical correlation length and spatio-temporal scaling laws. BB also 
proposed a Ginzburg criterion that delineates the region of validity of MCT, which breaks down in low dimensions. Still, 
the field theory language used in \cite{EPL} is not trivially related to the standard, liquid theory formulation of MCT 
\cite{Gotze1}. Indeed recent work has shown 
that the field theory is laden with subtleties  \cite{MR,BLA,CatesRamas}, in particular related to 
the Fluctuation-Dissipation relation. The aim of the present letter is twofold. First, we show how the results 
of BB may be recovered and extended to obtain testable, quantitative predictions on absolute dynamic length scales,
entirely within via the standard, projection-operator based MCT \cite{Gotze1}. Our detailed analysis predicts a 
remarkable scaling behavior that has implications for the geometry of dynamic heterogeneities.  
Second, our formulation generalizes MCT to spatially inhomogeneous situations.

In order to proceed, we consider an atomic fluid subject to an arbitrary external potential $U(\bx)$, such that the equilibrium averages 
(e.g. the static density $\rho(\bx)$) vary in space. This opens a route to an MCT analysis of various situations of 
experimental interest, such as liquids in pores \cite{Krakoviak}, close to a wall \cite{wall} or a free surface, 
or sedimentation effects. The relationship with the results of BB will be obtained using this inhomogeneous MCT formalism to 
compute the response of the dynamical structure factor to a localized external potential.
In the limit where the wave vector associated with the external field tends to zero, a connection to the 4-point 
correlator of BB emerges. The dynamical quantities of interest are the density fluctuations $\delta \rho$ and the currents
$J$, defined in Fourier space as $\delta\rho_{\bq} = \sum_{i=1}^{N}\e^{i\bq\cdot\br_i} - \phi(\bq)$ and $J_{\bq} = 
\sum_{i=1}^{N}{\hq\cdot\bp_i}\e^{i\bq\cdot\br_i}/m$, where 
$\phi(\bq) \equiv {1}/{N}\lgle \sum_{i=1}^{N}\e^{i\bq\cdot\br_{i}}\rgle$.
Following standard procedures based on the Mori-Zwanzig formalism \cite{Gotze1,Dave}, 
one can establish the following {\it exact} equation of motion for the dynamic 
structure factor $F(\bq_1, \bq_2; t) = {1}/{N}\lgle \delta \rho_{\bq_1}(t)\delta \rho_{\bq_2}^{\ast}(t=0)\rgle$:
\begin{equation}\label{IMCT}
\begin{aligned}
& \frac{\partial^2~}{\partial t^2}F(\bq_1,\bq_2;t)
+\int d\bq_1' \Omega^2(\bq_1,\bq_1^{\prime})F(\bq_1^{\prime},\bq_2;t) +
\\
& \int d\bq_1'\int_{0}^{t}\!\!\dd t'~ M(\bq_1,\bq_1^{\prime};t-t')\pdif{~}{t'}F(\bq_1^{\prime},\bq_2;t') =0,
\end{aligned}
\end{equation}
where $\Omega^2(\bq_1,\bq_1^{\prime}) \equiv \frac{\kb T}{m}\bq_1\cdot\bq_1^{\prime} \phi(\bq_1-\bq^{\prime}_1)
S^{-1}(\bq_1,\bq_1^{\prime})$ (with $S^{-1}(\bq_1,\bq_2)$ the inverse operator of $F(\bq_1,\bq_2,t=0)$), and $M(\bq_1,\bq_2; t)$ is 
a memory kernel which can be expressed in terms of the fluctuating part of the force. Factorization approximations,
analogous to those used in standard MCT, 
reduce this memory kernel to two-body correlation functions and yield an inhomogeneous mode-coupling theory (IMCT). 
The general IMCT equations are rather cumbersome and will be presented elsewhere \cite{BBMRlong}. 
In the limit $U({\bf x}) \rightarrow 0$, they reduce to the standard MCT equations. In order to obtain 
somewhat tractable expressions, one can consider {\it weakly} inhomogeneous situations $U(\bx) \ll \kb T$, such that one can expand all quantities
to first order in the perturbation. The aim is to compute the sensitivity of the dynamical structure factor to 
a small perturbation of arbitrary spatial structure, in particular localized perturbations, which we can always decompose in 
Fourier modes:
$
\frac{\delta F(\bx,\by,t)}{\delta U(\bz)}\left.\right|_{U=0}=\int d\bq_1d\bq_0 e^{-i\bq_1\cdot(\bx-\by)+i\bq_0\cdot(\by-\bz)}
\chi_{\bq_0}(\bq_1,t)$, where 
$\chi_{\bq_0}(\bq_1,t) \propto \frac{\delta F(\bq_1,\bq_0+\bq_1,t)}{\delta U(\bq_0)}\left.\right|_{U=0}$
is the response of the dynamical structure factor to a static external perturbation in Fourier space. 
For a perturbation localized at the origin, $U(\bx)=U_0\delta(\bx)$, one finds 
$\delta F(\bq_1,\by,t)=\int d\bq_0e^{i\bq_0\cdot\by}\chi_{\bq_0}(\bq_1,t)$. 
This susceptibility is akin (although not exactly related) to a three-point density correlation 
function in the absence of the perturbation. Although quite different from 
the four-point functions considered previously in the literature, $\chi_{\bq_0}(\bq_1,t)$ is expected to reveal the 
existence of a dynamical correlation length of the homogeneous liquid (see \cite{Science} for the particular case $\bq_0=0$). 
Indeed, $\chi_{\bq_0}(\bq_1,t)$ measures the influence of a density fluctuation at a given point in space on the 
dynamics elsewhere. Within MCT, this three-point correlation turns out to have exactly the same critical 
behaviour as the one obtained by BB. As shown below, the deep underlying theoretical reason for such a 
coincidence is that a certain linear operator becomes critical at the transition (see \cite{BBBKMR} for a diagrammatic explanation). 
Differentiating Eq. (\ref{IMCT}) with 
respect to $U(\bq_0)$, the final equation for the susceptibility $\chi_{\bq_0}(\bq_1,t)$ reads: 
\begin{widetext}
\begin{equation}\label{chi3}
\begin{aligned}
&
\frac{\partial^2 \chi_{\bq_0}(\bq_1,t)}{\partial t^2}
+ \frac{\kb T q_1^2}{m S(q_1)} \chi_{\bq_0}(\bq_1,t)
+\int_{0}^{t}\!\!\dd t'~ M_{0}(q_1,t-t')
  \pdif{\chi_{\bq_0}(\bq_1,t')}{t'} +\int_{0}^{t}\!\!\dd t'~ \frac{\kb T \rho q_1}{m |\bq_1+\bq_0|}
\\
&
\int\!\!\frac{\dd\bk}{(2\pi)^3} v(\bq_1;\bk,\bq_1-\bk)
v(\bq_1+\bq_0;\bq_1-\bk,\bq_0+\bk) \chi_{\bq_0}(\bk,t-t')F_0(|\bq_1-\bk|,t-t') \pdif{F_0(|\bq_1+\bq_0|,t')}{t'}
={\cal S}_{\bq_0}(\bq_1,t)
\end{aligned}
\end{equation}
\end{widetext}
where $v$ is the usual MCT vertex $v(\bq;\bk_1,\bk_2)= \hq\cdot\bk_1 nc(k_1)
+\hq\cdot\bk_2nc(k_2)$; $M_0$ is the MCT memory kernel: $M_{0}(q,t) = {\kb T \rho}/{2m}\int {\dd\bk}/{(2\pi)^3}
v^2(\bq;\bk,\bq-\bk)F_0(k,t)F_0(|\bk-\bq|,t)$ \cite{Dave}. The source term ${\cal S}_{\bq_0}(\bq_1,t)$, whose precise form will be presented elsewhere 
\cite{BBMRlong}, depends on 
$F_0(\bk,t'<t)$ and on static four and five point density correlations; the value of the dynamic length scale and the critical properties
of $\chi_{\bq_0}$ are however independent of the precise form of this source term. The above equation is of the type ${\cal L}_{\bq_0}\chi_{\bq_0}={\cal S}_{\bq_0}$, where ${\cal L}_{\bq_0}$ is a certain 
linear operator, the structure of which -- in particular its smallest eigenvalue -- contains the information we want
to extract. One should first note that in the limit $\bq_0 = 0$, the operator ${\cal L}_0$ simply encodes the change of 
the dynamic structure factor when the coupling 
constant (i.e. the density or the temperature) is shifted uniformly in space. 
This remark allows one to compute $\chi_0(\bk,t)$ from standard MCT results in 
the $\beta$ and $\alpha$ regimes:
\begin{equation}
\chi_0(\bk,t)= 
\left\{
\begin{aligned}
&
\frac{S(k)h(k)}{\sqrt{\varepsilon}}g_{\beta}\left(q_0^2=0,\frac{t}{\tau_\beta}\right); \quad
\tau_{\beta}= \varepsilon^{-1/2a},
\\
&
\frac{1}{\varepsilon}g_{\alpha,k}\left(\frac{t}{\tau_\alpha}\right); \quad
\tau_{\alpha}= \varepsilon^{-1/2a-1/2b},
\end{aligned}
\right.
\end{equation}
where we use standard MCT notations \cite{Gotze1,Dave} ($S(k)$ is the structure factor and 
$h(k)$ is the critical amplitude and $\epsilon$
is the distance, in the liquid phase, between the coupling constant and 
its critical value). The behavior of the scaling 
functions at large and small argument can be found directly by analyzing Eq. (\ref{chi3}) or by scaling: 
in the early $\beta$ regime $u = t/\tau_\beta\to 0$ the 
$\varepsilon$ dependence should drop out, hence $g_{\beta}(0,u)\propto u^a$. The matching between $\alpha$ and $\beta$ 
regime implies $g_{\beta}(0,u)\propto u^b$ at large $u$, whereas $g_{\alpha,k}(u)\propto S(k)h(k)u^b$ at small
$u$.
How is the behaviour modified for small modulations with $\bq_0 \neq 0$? The derivation is simple for 
$ \chi_{\bq_0}(\bq_1,\infty)$ in the glass phase, where straightforward manipulations of Eq. (\ref{chi3}) allows one to 
show that it satisfies the matrix equation $(I-M)\cdot \chi_{\bq_0}(\bq_1,\infty)=s_{\bq_0}$ where $s_{\bq_0}$ is a 
source term that is regular and of order one in the limit $q_0\rightarrow 0$ and:
\begin{equation}
\begin{aligned}
&
M_{\bq_0}(\bq_1, \bq_2)=\frac{\rho G(\bq_1)G(\bq_0+\bq_1)S(\bq_1-\bq_2)}{(2\pi)^3q_1 |\bq_1+\bq_0|}
\\
&
f_{\bq_1-\bq_2}
v(\bq_1;\bq_2,\bq_1-\bq_2)v(\bq_1+\bq_0;\bq_1-\bq_2,\bq_0+\bq_2),
\end{aligned}
\end{equation}
where $G(\bq_1)=S(\bq_1)(1-f_{\bq_1})$ and $f_\bq$ is the non-ergodic parameter. Interestingly, the matrix $M$ is 
{\it exactly} the same as the one obtained
from the resummation of the ladder diagrams in the field theoretical framework of BB. 
Note also that the source $s_{\bq_0}$ is irrelevant provided it is not orthogonal to the 
lowest eigenvector of $M$.  For $q_0=0$ and $\varepsilon < 0$, the largest eigenvalue of $M$ is, as shown by G{\"o}tze, 
$\lambda=1-O(\sqrt{-\varepsilon})$ \cite{Gotze85} and its right eigenvector is $S(k)h(k)$. The correction $\delta \lambda$ 
to this eigenvalue at $q_0 \to 0$ can be computed by perturbation theory. By symmetry, one expects that in general 
$\delta \lambda = -\Gamma q_0^2$, where $\Gamma$ is a certain coefficient, leading to $\chi_{\bq_0}(\bk,\infty) \sim
S(k)h(k)/(\sqrt{|\varepsilon|} + \Gamma q_0^2)$ \cite{EPL}. In the schematic limit where $S(\bq)$ is 
sharply peaked around $q=Q$, with a small width $\Delta Q$, one can compute $\Gamma$ exactly; one finds that 
$\Gamma$ is positive and $\propto \Delta Q^{-2}$.
For a more realistic shape of $S(q)$, e.g. 
hard sphere structure computed within the Percus-Yevick approximation at 
the mode-coupling density $\phi_{c}=0.515$, we have determined $\delta 
\lambda$ numerically to extract $\Gamma=0.072 \sigma^{-2}$.  For more 
realistic hard-sphere structure factors $\Gamma$ may be as large as $\Gamma=0.3 \sigma^{-2}$.
We have not been able to show in full generality that $\Gamma$ should always be positive. A negative $\Gamma$ would predict a 
remarkable `modulated' glass transition, with the non-ergodic factor displaying periodic oscillations in space \cite{ft4}. 

The analysis of the $\beta$ and $\alpha$ regime in the liquid phase 
is more involved. As previously, the operator that becomes critical at the transition turns out to be the same 
as the one considered in BB (and whose inversion leads to ladder diagrams). We find \cite{BBMRlong} that 
$(I-M_c)\cdot \chi_{\bq_0}(t\varepsilon^{1/2a})=s^{\prime}_{\bq_0}(t\varepsilon^{1/2a})$, where $M_c$ is the matrix $M$
at the transition and the new source term 
is of order one in the limit $q_0\rightarrow 0$. As a consequence one finds\cite{BBMRlong}:  
\[
\chi_{\bq_0}(\bk,t)=\frac{1}{\sqrt{\varepsilon}+\Gamma q_0^2}S(k)h(k) \, g_{\beta}\left(\frac{q_0^2}{\sqrt{\varepsilon}},
t\varepsilon^{1/2a}\right)
\]
where $g_{\beta}$ satisfies $q_0^2 g_{\beta}(q_0^2/\sqrt{\varepsilon},t\varepsilon^{1/2a})=
(\sqrt{\varepsilon}+\Gamma q_0^2)\langle l|s^{\prime}_{\bq_0}(t\varepsilon^{1/2a})\rangle$ with $\langle l|$ is the left eigenvector conjugated to 
$S(k)h(k)$. 
\begin{figure}
\begin{center}
\psfig{file=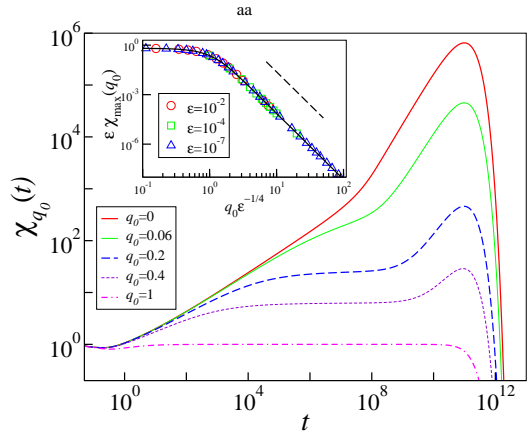,width=5.5cm,angle=270}
\caption{
Numerical solution of the schematic IMCT equations for $T > T_c$, where all $k$ dependence is 
neglected (see \cite{BBMRlong} for details). Main plot: $\chi_{\bq_0}(t)$ for different $\bq_0$ as
a function of time, and for $\varepsilon=10^{-6}$. From top to bottom: $q_0=0,0.06,0.2,0.4,1.$. 
Note that the {\it shape} of $\chi_{\bq_0}(t)$ in the $\alpha$ 
regime is independent of $\bq_0$, as predicted by Eq. (\ref{latebeta}). We have in fact checked that the 
predicted scaling is very well obeyed in that region. Inset: $\varepsilon \chi_{\max}(\bq_0) \equiv \varepsilon 
\chi_{\bq_0}(t=\tau_\alpha)$ as a function of $q_0 \varepsilon^{-1/4}$ in log-log, for different $q_0$'s and $\varepsilon$'s. Note the
$q_0^{-4}$ behaviour for large $q_0 \varepsilon^{-1/4}$, as indicated by the dashed line.}
\end{center}
\end{figure}
The analysis of $\chi_{\bq_0}(\bk,t)$ in the $\alpha$-relaxation regime turns out to be more subtle than anticipated in BB 
and will be detailed elsewhere \cite{BBMRlong}. We have established that for small $q_0$ and fixed $\varepsilon$, 
\be \label{latebeta}
\chi_{\bq_0}(\bk,t) =\frac{\Xi(\Gamma q_0^2/\sqrt{\varepsilon})}{\sqrt{\varepsilon} (\sqrt{\varepsilon} + \Gamma q_0^2)} 
\, g_{\alpha,k}\left(\frac{t}{\tau_\alpha}\right),
\ee
with $\Xi$ a certain regular function with $\Xi(0) \neq 0$ and $\Xi(v \gg 1) \sim 1/v$ such that $\chi_{\bq_0}$ behaves as $q_0^{-4}$ for large $q_0$, independently of $\varepsilon$. Also, $g_{\alpha,k}(u \ll 1) = S(k)h(k)u^b $, as to match the 
$\beta$ regime, and 
$g_{\alpha}(u \gg 1,k) \to 0$. All those analytical results are in full agreement with the numerical solution of the 
schematic version of IMCT equations (see Fig. 1).
Note that the scaling variable is still $q_0^2 \varepsilon^{-1/2}$ in that regime, rather than $q_0^2 \varepsilon^{-1}$
as surmised in BB \cite{EPL}. The physical consequence of the above analysis is the existence of a unique diverging 
length scale $\xi \sim \sqrt{\Gamma} |\varepsilon|^{-1/4}$ that rules the response of the system to a 
space-inhomogeneous perturbation and hence of the spatial dynamic correlations. 
The analysis of the early $\beta$ regime where $t \ll |\varepsilon|^{-1/2a}$ \cite{BBMRlong} shows that this length in fact 
first increases as $t^{a/2}$ and then saturates at $\xi$. Furthermore, Eq. (\ref{latebeta}) indicates that 
although the integrated dynamic correlation $\chi_{\bq_0=0}(\bk,t)$ increases in the $\alpha$ regime as 
$\varepsilon^{(b-a)/2a} t^b$ (from 
$\varepsilon^{-1/2}$ for $t=\tau_\beta$ to $\varepsilon^{-1}$ for $t=\tau_\alpha$) the dynamic length scale itself 
remains fixed at $\xi$. Interestingly, this suggests that while keeping a fixed extension $\xi$, the (fractal) geometrical 
structures carrying the dynamic correlations significantly `fatten' \cite{ftnfrac} between $\tau_\beta$ (where the 
structures could correspond
to the strings reported in recent simulations \cite{Glotzer,Heuer}) and $\tau_\alpha$, where more compact 
structures are expected, as indeed suggested by the results of \cite{Kob}. For $\tau_\beta \ll t \ll \tau_\alpha$, we 
expect a cross-over between dense and dilute structures at a new, time dependent crossover length \cite{BBMRlong}.

Starting from the general IMCT equation Eq. (\ref{IMCT}), one could have chosen to follow a slightly different path and 
only assume that the length scale $\ell$ of the imposed inhomogeneities is large, and perform a gradient expansion to
order $\ell^{-2}$ to obtain an equation on the space dependent structure factor, $F(\bq,\br,t)$.
This space-dependent Ginzburg-Landau like MCT equation has one part that is identical to standard MCT equation (with
space dependent coefficients) plus non linear contributions containing a $\nabla^2 F$ term and, interestingly, a
Burgers non-linear term of the form $(\nabla F)^2$ \cite{BBMRlong}. When inhomogeneities are small, 
$F(\bq,\br,t)=F_0(\bq,t)+\chi(\bq,\br,t)$ with $\chi \ll F_0$, these equations become identical to Eq. (\ref{chi3}) above. 
In the schematic limit where all wave-vector dependencies drop off, our equation 
coincides with the gradient expansion obtained by Franz for the p-spin glass model in the Kac limit \cite{Franz}. \\
In summary, we have shown how to extend the standard framework of MCT to inhomogeneous situations. 
This allows us to compute the response of the dynamical structure factor to spatial perturbations.
The case of a localized perturbation at the origin is particularly interesting:  
it shows directly that the dynamical structure factor is affected on a dynamic length scale 
$\xi$ that diverges as $|\varepsilon|^{-\nu}$ as the Mode-Coupling transition is approached, 
with $\nu=1/4$ and a prefactor that can be computed numerically.
This length scale governs both the $\beta$ and $\alpha$ relaxation regimes, showing that the 
standard interpretation of the $\beta$-regime as the vibrations of particles trapped in  
independent cages formed by nearest neighbors is somewhat misleading: as the MCT transition 
is approached, $\xi$ grows and the `cages' become more and more collective.
Our results suggest an interesting scenario where the geometrical structures 
responsible for dynamic fluctuations thicken with time. Note that $\xi$, as in ordinary critical phenomena,
only diverges at $T_c$, reflecting the 
critical fragility of the system right at the transition. It is therefore clearly 
distinct from the diverging viscous length 
$\sqrt{\eta \tau_\alpha}$ which sets the scale below which the liquid sustains shear waves \cite{Das}, which  
is infinite in the whole glass phase. Our IMCT equations
should be very useful to investigate inherently inhomogeneous physical situations.  Also, spatial critical fluctuations
are expected to play a major role in low dimensions, as for usual critical phenomena. These should lead to non trivial 
values of critical exponents (such as $\nu$) and be involved in the breakdown of the Stokes-Einstein relation between viscosity and
diffusion \cite{BBdecoup}. However, these critical fluctuations should also interfere with activated `droplet' fluctuations 
that are expected to smear out the MCT transition in finite dimensions \cite{BBChem}. The details of the way these
two type of phenomena interact and lead to the observed crossover between an MCT-like regime and an activated 
regime in supercooled liquids is, in our opinion, one of the most crucial open theoretical questions. Mixing the present 
framework with the extended Mode-Coupling scheme recently proposed in \cite{eMCT} might be a promising path. Finally, 
we want to stress that generalized susceptibilities such as $\chi_{\bq_0}(\bk,t)$ offer a direct way to study 
correlations between dynamics and local structural fluctuations \cite{Science}.  They provide a complementary and 
more direct physical information than the 4-point correlations studied in \cite{Parisilength,Glotzer,Berthier}. 
We notice that $\chi_{\bq_0}(\bk,t)$ can be measured using state of the art molecular or Brownian dynamics simulation.
Experimentally, it could be accessible in colloids by use of an optical tweezer array imposing a periodic dielectric force 
on the particles.

We warmly thank L. Berthier, S. Franz, W. Kob, V. Krakoviack, G. Tarjus, M. Wyart and F. Zamponi.
GB is partially supported by EU contract HPRN-CT-2202-00307 (DYGLAGEMEM) and DRR by grants 
NSF CHE-0505939 and NSF DMR-0403997.


\begin{thebibliography}{9}
\bibitem{ftnfrac} Our results suggest that in high dimensions, the fractal dimension of the dynamic clusters increases 
from $d_f=2$ in the $\beta$ regime to $d_f=4$ in the $\alpha$ regime.
\end{thebibliography}
\end{document}